\documentclass[10pt]{article}
\usepackage[latin9]{inputenc}
\usepackage{geometry}
\usepackage{units}
\usepackage{mathtools}
\usepackage{amsmath}
\usepackage{amsthm}
\usepackage{amssymb}
\usepackage{graphicx}

\makeatletter

\newcommand{\noun}[1]{\textsc{#1}}
\providecommand{\tabularnewline}{\\}

\numberwithin{equation}{section}
\numberwithin{figure}{section}

\usepackage{amsfonts}
\usepackage{amscd}
\usepackage{array}\usepackage{dsfont}\usepackage{graphicx}\usepackage{epsfig}
\usepackage{array}\usepackage{latexsym}\usepackage{xspace}\usepackage{comment}

\usepackage{graphicx}

\makeatother

\begin{document}
\title{\noun{Physics with non-unital algebras?} \\
\emph{An invitation to the Okubo algebra}}
\author{Alessio Marrani$^{\dagger}$, Daniele Corradetti$^{*}$, Francesco
Zucconi$^{\ddagger}$}
\maketitle
\begin{abstract}
This paper presents some preliminary discussion on the possible relevance of
the Okubonions, i.e. the real Okubo algebra $\mathcal{O}$, in quantum
chromodynamics (QCD). The Okubo algebra lacks a unit element and sits in the
adjoint representation of its automorphism group $\text{SU}_{\mathcal{O}}$, thus
being fundamentally different from the better-known octonions $\mathbb{O}$.
While these latter may represent quarks (and color singlets), the Okubonions
are conjectured to represent the gluons, i.e. the gauge bosons of the QCD $%
\text{SU}(3)$ color symmetry. However, it is shown that the $\text{SU}(3)$ groups pertaining to Okubonions and octonions are distinct and inequivalent
subgroups of $Spin(8)$ that share no common $\text{SU}(2)$ subgroup. The unusual
properties of Okubonions may be related to peculiar QCD phenomena like
asymptotic freedom and color confinement, though the actual mechanisms remain to
be investigated.
\end{abstract}

\tableofcontents

\newpage

\section*{Introduction}

\textit{Quantum chromodynamics} (QCD) is the theory of the strong
interaction between \textit{quarks} mediated by \textit{gluons}, and it is
an important part of the \textit{Standard Model of particle physics}
(henceforth abbreviated as SM); for a nice recent introduction and a list of
Refs., see e.g. \cite{QCD} and \cite{QCD-Wilczek}. QCD is a
non-Abelian (Yang-Mills) gauge theory, with exact symmetry group $\text{SU}%
\left( 3\right) $, whose charge is named \emph{color}, such that the QCD
gauge group is notated as SU$(3)_{\text{color}}$. \textit{Quarks} are
fundamental, massive fermions, with spin $1/2$, carrying a color charge in
the fundamental representation $\mathbf{3}$ of $\text{SU}\left( 3\right) _{%
\text{color}}$, along with a fractional electric charge (either $-1/3$ or $%
+2/3$). They participate in the weak interactions as part of the weak
isospin doublets, and come in six flavours, denoted as $u$ (up)$,$ $d$ (down)%
$,$ $c$ (charm)$,$ $s$ (strange)$,$ $t$ (top)$,$ $b$ (bottom); each type of
quark has a corresponding antiquark, whose (color and electric) charges are
exactly opposite : e.g., antiquarks transform in the conjugate $\text{SU}%
\left( 3\right) _{\text{color}}$ -representation to quarks, denoted $%
\overline{\mathbf{3}}$. On the other hand, gluons are fundamental, massless
bosons, with spin $1$, also carrying color charges, but lying in the adjoint
representation $\mathbf{8}$ of $\text{SU}\left( 3\right) _{\text{color}}$ ;
they have no electric charge, and do not participate in the weak
interactions.

QCD exhibits a number of counterintuitive, `weird' features, such as \textit{%
color confinement} \cite{Greensite} and \textit{asymptotic freedom} \cite%
{GW, Poli}. Thus, the following question arises quite naturally : \textit{%
can some of the features exhibited by QCD be explained by modeling the
corresponding fundamental particles, namely quarks and gluons, in terms of
some algebra?}

In this work, we will not attempt at answering this intriguing question in a
complete and satisfactory way, but we will rather aim at suggesting a
further ingredient - which exhibits some quite unique features - within the
quest for an algebraic modeling of QCD: the 8-dimensional division algebra
of \textit{Okubonions} (\emph{aka} the Okubo algebra) $\mathcal{O}$. This
algebra is non-alternative and lacks the unit element; moreover, it has the
quite unique feature to sit in the adjoint representation of its
automorphism group. It was formally discovered by Petersson in the late '60s,
and then independently rediscovered by Okubo almost ten years later \cite{Ok1,Ok2}
(see also \cite{Ok-book} for a review and a list of Refs.). However,
Okubonions did not receive much attention by mathematicians (and physicists
as well) until quite recently, when they were investigated by Elduque and
Myung in a series of works \cite{El90}-\nocite{El91,El93,El94,El96,El15}\cite{El21}.

In the present paper, we will proceed by contrasting the properties of the
(real) Okubo algebra $\mathcal{O}$ (which is non-unital and non-alternative)
to the properties of real (division) octonions $\mathbb{O}$ (\emph{aka}
Cayley numbers). This latter is a 8-dimensional non-associative (but
alternative and unital) Hurwitz algebra, which has fascinated mathematicians
and physicists for decades, and which has recently witnessed a renewed
interest and application within the algebraic modeling of SM interactions
(see e.g. \cite{Baez-O} for a nice review and a list of Refs.).
%we will put forward a quite obvious, yet - to the best of our knowledge -
%novel and intriguing, physical interpretation of the Okubonions $\mathcal{O}$
%within QCD, which however, curiously, is complementary to the physical
%interpretation of octonions $\mathbb{O}$ in the same framework.
\bigskip

The plan of the paper is as follows.

In Sec. 1 we review the main properties of all the three 8-dimensional division
and composition algebras (over $\mathbb{R}$): the octonions $\mathbb{O}$,
the para-octonions $p\mathbb{O}$ (not to be confused with the
split-octonions $\mathbb{O}_{s}$), and the Okubonions $\mathcal{O}$. In Sec.
2 we recall the interpretation of $\mathbb{O}$ in QCD. Then, in Sec. 3 we
put forward an interpretation of $\mathcal{O}$ within the same framework,
whereas in Sec. 4 we prove a crucial difference between the $\text{SU}\left(
3\right) $ symmetries pertaining to $\mathbb{O}$ and $\mathcal{O}$. Finally,
some conclusions are drawn in Sec. 5.\medskip

Before starting, a few general remarks. Since we work over $\mathbb{R}$, we
will call the Cayley algebra over the field of real as the \textquotedblleft
octonions\textquotedblright\ $\mathbb{O}$ (rigorously identified by the
triplet $\left( \mathbb{O},\cdot ,n\right) $), and the Okubo algebra $%
\mathcal{O}$ (rigorously identified by the triplet $\left( \mathbb{O},\ast
,n\right) $) over the reals as \textquotedblleft
Okubonions\textquotedblright\ (or simply \textquotedblleft Okubo
algebra\textquotedblright ). Moreover, we will be using the usual
physicists' notation of irreducible representations (irreprs.) of Lie groups
and algebras, which identifies an irrepr. with its dimension in bold, along
with some further marks if needed.

\section{The three 8-dimensional, real, division, composition algebras : $%
\mathbb{O}$, $p\mathbb{O}$, and $\mathcal{O}$}

An algebra $A$ is a vector space endowed with a bilinear product. The
specific properties of such bilinear product (which we will denote by $\cdot
$) lead to various classifications: an algebra $A$ is said to be\emph{\
commutative} if $x\cdot y=y\cdot x$ for every $x,y\in A$; is \emph{%
associative} if satisfies $x\cdot \left( y\cdot z\right) =\left( x\cdot
y\right) \cdot z$; is \emph{alternative} if $x\cdot \left( y\cdot y\right)
=\left( x\cdot y\right) \cdot y$; and finally, \emph{flexible} if $x\cdot
\left( y\cdot x\right) =\left( x\cdot y\right) \cdot x$. If the
algebra is also equipped with a norm $n$, then it is a normed
algebra. Moreover, if the norm respects the multiplicative structure, i.e., $%
n\left( x\cdot y\right) =n\left( x\right) n\left( y\right) $, the algebra is
called a \emph{composition algebra}. Thus, a composition algebra is fully
identified by the triple $\left( A,\cdot ,n\right) $.

Composition algebras split into \emph{unital}, i.e. where exists a unit
element $1$ such that $x\cdot 1=1\cdot x=x$, \emph{para-unital}, i.e. where
exists an involution $x\longrightarrow \overline{x}$ and an element called
para-unit $\boldsymbol{1}$ such that $x\cdot \boldsymbol{1}=\boldsymbol{1}%
\cdot x=\overline{x}$, and \emph{non-unital}, namely algebras that do not
possess neither a unit element nor a para-unit element. %\begin{table}
A complete classification is provided by the Generalized Hurwitz Theorem
\cite{El21}, which states that only sixteen composition algebras exist (e.g.on $\mathbb{R}$)
: seven are unital and called \emph{Hurwitz algebras},
including the division algebras $\mathbb{R},\mathbb{C},\mathbb{H},\mathbb{O}$
along with their split companions $\mathbb{C}_{s},\mathbb{H}_{s},\mathbb{O}%
_{s}$; another seven are para-unital, closely related to the Hurwitz
algebras and termed \emph{para-Hurwitz algebras}; finally, there are two
composition algebras, one division and one split, that are both non-unital
and 8-dimensional, known as the \emph{Okubo algebras} $\mathcal{O}$ and $%
\mathcal{O}_{s}$ \cite{El91}.

Out of such sixteen composition algebras, in the present treatment we will be interested only in the 8-dimensional division
ones, namely in the following three (mutually
non-isomorphic) algebras : the octonions $\mathbb{O}$ (which are alternative
and unital), the para-octonions $p\mathbb{O}$ (non-alternative and
para-unital) and the Okubo algebra $\mathcal{O}$ (non-alternative and
non-unital). Their properties are summarized in Table \ref%
{tab:Synoptic-table-of}.
\begin{table}
\centering{}%
\begin{tabular}{|c|c|c|c|}
\hline
Property & $\mathbb{O}$ & $p\mathbb{O}$ & $\mathcal{O}$\tabularnewline
\hline
\hline
Unital & Yes & No & No\tabularnewline
\hline
Para-unital & Yes & Yes & No\tabularnewline
\hline
Alternative & Yes & No & No\tabularnewline
\hline
Flexible & Yes & Yes & Yes\tabularnewline
\hline
Composition & Yes & Yes & Yes\tabularnewline
\hline
Automorphism & $\text{G}_{2}$ & $\text{G}_{2}$ & $\text{SU(3)}$\tabularnewline
\hline
\end{tabular}\caption{\label{tab:Synoptic-table-of}Synoptic table of the algebraic properties
of octonions $\mathbb{O}$, para-octonions $p\mathbb{O}$ and the real
Okubo algebra $\mathcal{O}$.}
\end{table}

\subsection{The algebra of octonions $\mathbb{O\equiv }\left( \mathbb{O}%
,\cdot ,n\right) $}

\begin{figure}[h]
\begin{centering}
\includegraphics[scale=0.8]{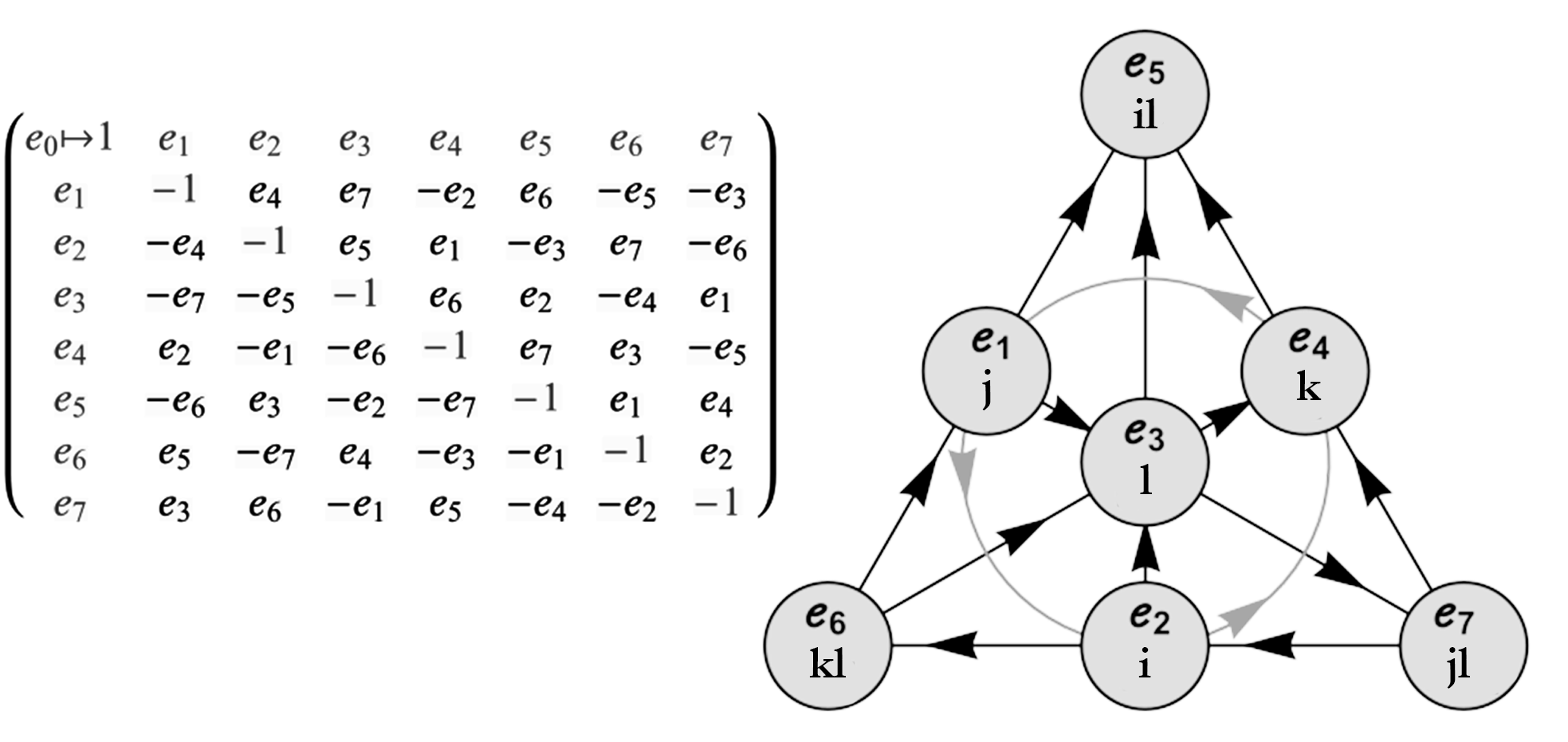}
\par\end{centering}
\caption{\emph{On the left}: octonionic multiplication tables for the basis $%
\left\{ e_{0}=1,e_{1},e_{2},e_{3},e_{4},e_{5},e_{6},e_{7}\right\} $. \emph{%
On the right}: a mnemonic representation on the Fano plane of the same
octonionic multiplication rule with the equivalence with the Dickson
notation $\left\{ 1,\text{i},\text{j},\text{k},\text{l},\text{il},\text{jl},%
\text{kl}\right\} $ according to \protect\cite{Co46}.}
\label{fig:Fano-Plane}
\end{figure}

The \textit{octonions} $\left( \mathbb{O},\cdot ,n\right) $ are a Hurwitz
algebra, and they can be defined as the 8-dimensional real vector space with
basis $\left\{ e_{0}=1,e_{1},...,e_{7}\right\} $, endowed with a
bilinear product $\cdot $ encoded through the \emph{Fano plane} and
explained in Fig. \ref{fig:Fano-Plane}. The resulting algebra is
non-associative and non-commutative, but alternative (and thus flexible).

Given an element $x\in \mathbb{O}$ with decomposition
$x=x_{0}e_{0}+x_{1}e_{1}+...+x_{7}e_{7}$, one can define its conjugate
element as
$\overline{x}:=x_{0}e_{0}-x_{1}e_{1}-...-x_{7}e_{7}$; as in
any unital composition algebra, $\overline{x}$ is the image of $x$ under an
order-two homomorphism given by a canonical involution, named \emph{%
conjugation}. The norm $n$ is the obvious Euclidean one, defined by
\begin{equation}
n\left( x\right)
=x_{0}^{2}+x_{1}^{2}+x_{2}^{2}+x_{3}^{2}+x_{4}^{2}+x_{5}^{2}+x_{6}^{2}+x_{7}^{2},
\label{eq:octonionic norm}
\end{equation}%
and, therefore,
\begin{equation}
n\left( x\right) =x\cdot \overline{x}=\overline{x}\cdot x,
\label{eq:n(x)=00003Dxcx}
\end{equation}%
as it holds for every Hurwitz algebra. (\ref{eq:octonionic norm}) yields
that $n\left( x\right) =0\Leftrightarrow x=0$ and thus the inverse of a
non-zero element of the octonions is easily found as $x^{-1}=\overline{x}%
/n\left( x\right) .$ Also, the polarization of (\ref{eq:n(x)=00003Dxcx})
yields the definition of the octonionic inner product as $\left\langle
x,y\right\rangle =x\cdot \overline{y}+y\cdot \overline{x},$ so that $%
\left\langle x,x\right\rangle =2n\left( x\right) $. Note that $\mathbb{O}$ is a
division algebra, since if $\left\langle x,y\right\rangle =0$, then either $x$ or $y$ are zero.

By exploiting the unitality of $\mathbb{O}$, the octonionic conjugation can
equivalently be defined using the orthogonal projection on the unit element $1$ as
\begin{equation}
x\mapsto \overline{x}:=\left\langle x,1\right\rangle 1-x.
\label{eq:coniugazione}
\end{equation}%
This canonical involution has the distinctive property of being an
antihomomorphism with respect to the product, i.e. $\overline{x\cdot y}=%
\overline{y}\cdot \overline{x}$.

\subsection{The algebra of para-octonions $p\mathbb{O\equiv }\left( \mathbb{O}%
,\bullet ,n\right) $}

Starting from the Hurwitz algebra of octonions $\left( \mathbb{O},\cdot
,n\right) $, a para-Hurwitz algebra can be introduced by defining a new
product
\begin{equation}
x\bullet y=\overline{x}\cdot \overline{y},
\end{equation}%
for every $x,y\in \mathbb{O}$. The new algebra $\left( \mathbb{O},\bullet
,n\right) $ is again a composition algebra, in fact a para-Hurwitz algebra,
called \textit{para-octonions} and denoted with $p\mathbb{O}$ (not to be
confused with the algebra of split-octonions $\mathbb{O}_{s}$, which is a
Hurwitz algebra with zero divisors) \cite{Knus,El21}. Para-octonions do not
have a unit, but only a para-unit, i.e. $\boldsymbol{1}\in p\mathbb{O}$ such that $%
\boldsymbol{1}\bullet x=x\bullet \boldsymbol{1}=\overline{x}.$ Note that $p\mathbb{O}$ is still a
division algebra, since if $x\bullet y=0,$ then either $\overline{x}$ or $\overline{y}$ are zero, thus implying that
either $x$ or $y$ are zero.

\subsection{The Okubo algebra $\mathcal{O}\equiv \left( \mathbb{O},\ast
,n\right) $ : Petersson's and Okubo's realizations}

$\mathbb{O}$ also admits an order-three involutive automorphism $\tau $; considering again the basis $%
\left\{ e_{0}=1,e_{1},...,e_{7}\right\} $ of $\mathbb{O}$ as a vector space, $\tau $ can be defined as follows \cite{Knus}:
\begin{equation}
\begin{array}{ccc}
\tau \left(e_{k}\right) & =e_{k}, & k=0,1,3,7 \\
\tau \left(e_{2}\right) & =-\frac{1}{2}\left(e_{2}-\sqrt{3}%
e_{5}\right) , & \tau \left(e_{4}\right) =-\frac{1}{2}\left(
e_{4}-\sqrt{3}e_{6}\right) , \\
\tau \left(e_{5}\right) & =-\frac{1}{2}\left(e_{5}+\sqrt{3}%
e_{2}\right) , & \tau \left(e_{6}\right) =-\frac{1}{2}\left(
e_{6}+\sqrt{3}e_{4}\right) .%
\end{array}
\label{eq:Tau(Octonions)}
\end{equation}%
By starting from the Hurwitz algebra of octonions $\left( \mathbb{O},\cdot
,n\right) $, such a map can be exploited in order to obtain a new
(non-isomorphic) algebra (belonging to the class of Petersson algebras \cite%
{Petersson 1969}), whose product (denoted by $\ast $) is defined as
\begin{equation}
x\ast y=\tau \left( \overline{x}\right) \cdot \tau ^{2}\left( \overline{y}%
\right) ,
\end{equation}%
for every $x,y\in \mathbb{O}$. The new algebra $\left( \mathbb{O},\ast
,n\right) $ is again a composition algebra, called the \textit{Okubo algebra}
$\mathcal{O}$. This algebra does not have a unit nor a para-unit, but it has
idempotent elements. However, $\mathcal{O}$ is still a division algebra,
because if $x\ast y=\tau \left( \overline{x}\right) \cdot \tau ^{2}\left( \overline{y}%
\right) =0$, then either $\tau \left( \overline{x}\right) $ or $\tau ^{2}\left( \overline{%
y}\right) $ are zero, and recalling that $\tau $ is an automorphism, this implies that
either $x$ or $y$ are zero.

An independent realisation of $\mathcal{O}$, equivalent to the above realization
\textit{à la Petersson} based on $\tau $, was found by Okubo in the late '70s,
and it is based on a peculiar deformation of the Jordan
product. Following \cite{Ok1} and \cite{El91}, $\mathcal{O}$ can indeed
also be defined as the set of $3\times 3$ Hermitian traceless
matrices over the complex numbers $\mathbb{C}$, endowed with the product
\begin{equation}
x\ast y=\mu xy+\overline{\mu }yx-\frac{1}{3}\text{Tr}\left( xy\right) ,
\label{eq:product Ok}
\end{equation}%
where $\mu =\nicefrac{1}{6}\left( 3+\text{i}\sqrt{3}\right) $ and the
juxtaposition denotes the ordinary (non-commutative but associative) matrix product. It
should be remarked that $\ast $ is bilinear, but non-symmetric,
with its antisymmetric part proportional to Im$\mu =\frac{\sqrt{3}}{6}$;
 as anticipated, $\ast $ can be regarded as a deformation of the
Jordan product $x\circ y=\frac{1}{2}xy+\frac{1}{2}yx$, to which it
reduces by setting Im$\mu =0\Leftrightarrow \mu =\nicefrac{1}{2}$ and
neglecting the last term in the r.h.s. of (\ref{eq:product Ok}).
Nevertheless, the tracelessness property is not preserved under the Jordan
product $\circ $, and thus the additional term $-\nicefrac{1}{3}\text{Tr}%
\left( xy\right) $ is actually needed for the closure of the algebra. It is
amusing to notice that by simply setting $\text{Im}\mu =0$ (and retaining the
trace term in the r.h.s. of (\ref{eq:product Ok})), one retrieves the
traceless part $\mathfrak{J}_{3}\left( \mathbb{C}\right) _{0}$ of the
simple, cubic Jordan algebra $\mathfrak{J}_{3}\left( \mathbb{C}\right) $,
whose derivation Lie algebra is $\mathfrak{su}\left( 3\right) $.

Analyzing (\ref{eq:product Ok}), one can realize that the resulting
algebra $\mathcal{O}\equiv \left( \mathbb{O},\ast ,n\right) $, of real dimension $8$, is neither
unital, nor associative, nor alternative. However, $\mathcal{O}$ admits
idempotent elements, an example of which is
\begin{equation}
e=\left(
\begin{array}{ccc}
2 & 0 & 0 \\
0 & -1 & 0 \\
0 & 0 & -1%
\end{array}%
\right) ,  \label{eq:idemp}
\end{equation}%
which indeed satisfies $e\ast e=e$. By setting $e=e_{0}$ and defining
\begin{equation}
\begin{array}{ccc}
e_{1}=\sqrt{3}\left(
\begin{array}{ccc}
0 & 1 & 0 \\
1 & 0 & 0 \\
0 & 0 & 0%
\end{array}%
\right) , & e_{2}=\sqrt{3}\left(
\begin{array}{ccc}
0 & 0 & 1 \\
0 & 0 & 0 \\
1 & 0 & 0%
\end{array}%
\right) , & e_{3}=\sqrt{3}\left(
\begin{array}{ccc}
0 & 0 & 0 \\
0 & 0 & 1 \\
0 & 1 & 0%
\end{array}%
\right) , \\
e_{4}=\sqrt{3}\left(
\begin{array}{ccc}
1 & 0 & 0 \\
0 & -1 & 0 \\
0 & 0 & 0%
\end{array}%
\right) , & e_{5}=\sqrt{3}\left(
\begin{array}{ccc}
0 & -i & 0 \\
i & 0 & 0 \\
0 & 0 & 0%
\end{array}%
\right) , & e_{6}=\sqrt{3}\left(
\begin{array}{ccc}
0 & 0 & -i \\
0 & 0 & 0 \\
i & 0 & 0%
\end{array}%
\right) , \\
e_{7}=\sqrt{3}\left(
\begin{array}{ccc}
0 & 0 & 0 \\
0 & 0 & -i \\
0 & i & 0%
\end{array}%
\right) , &  &
\end{array}
\label{eq:definizione i ottonioniche}
\end{equation}%
one obtains an explicit realization of a basis for $\mathcal{O}$. In this
realization, the norm $n$ is defined through the matrix trace (denoted as Tr), namely%
\footnote{%
It can be proved that all the treatment is independent on the actual
explicit expression of the idempotent $e$.}
\begin{equation}
n\left( x\right) =\frac{1}{6}\text{Tr}\left( x^{2}\right) ,
\label{eq:Norm-Ok}
\end{equation}%
where $x^{2}\equiv xx$ (i.e., the square of $x$ under the usual matrix
product), for every $x\in \mathcal{O}$. It is then easy to see that the (quadratic) norm $n$ defined by (\ref%
{eq:Norm-Ok}) has Euclidean signature, and furthermore that it is associative and
composition over $\mathcal{O}$ itself.

\subsection{Transitions among $\mathbb{O}$, $p\mathbb{O}$ and $\mathcal{O}$}

As evident by glancing at their rigorous triplet notations, $\mathbb{O\equiv }%
\left( \mathbb{O},\cdot ,n\right) $, $p\mathbb{O\equiv }\left( \mathbb{O}%
,\bullet ,n\right) $ and $\mathcal{O}\equiv \left( \mathbb{O},\ast ,n\right)
$ are tightly related, as one can easily switch from one to the other by
simply changing the definition of the bilinear product over the vector space
of the algebra.

The transition $\left( \mathbb{O},\ast ,n\right) \longrightarrow \left(
\mathbb{O},\cdot ,n\right) $ can be realized by starting from $\mathcal{O}%
\equiv \left( \mathbb{O},\ast ,n\right) $ and defining the octonionic
product $\cdot $ as follows :
\begin{equation}
x\cdot y=\left( e\ast x\right) \ast \left( y\ast e\right) ,
\end{equation}%
where $x,y\in \mathcal{O}$ and $e$ is any idempotent element of $\mathcal{O}$
(e.g., given by (\ref{eq:idemp})). Since $e\ast e=e$ and $n\left( e\right)
=1 $, the element $e$ acts as a left and right identity, i.e.
\begin{align}
x\cdot e& =e\ast x\ast e=n\left( e\right) x=x, \\
e\cdot x& =e\ast x\ast e=n\left( e\right) x=x.
\end{align}%
Moreover, since $\mathcal{O}$ is composition, the same norm $n$ enjoys the
relation
\begin{equation}
n\left( x\cdot y\right) =n\left( \left( e\ast x\right) \ast \left( y\ast
e\right) \right) =n\left( x\right) n\left( y\right) ,  \label{tthis}
\end{equation}%
which means that $\left( \mathcal{O},\cdot ,n\right) $ is a unital
composition algebra of real dimension $8$. Since it is also a division
algebra, then it must be isomorphic to that of octonions $\mathbb{O}$, as
noted by Okubo himself \cite{Ok1,Ok2} : $\left( \mathcal{O},\cdot ,n\right)
\simeq \mathbb{O}$. On the other hand, the treatment of Okubonions given above explains how $\left(
\mathbb{O},\cdot ,n\right) \longrightarrow \left( \mathbb{O},\ast ,n\right) $
can be realized (actually, in a twofold way).

The scenario with para-octonions $p\mathbb{O}$ is also straightforward. The
transition $\left( \mathbb{O},\cdot ,n\right) \longrightarrow \left( \mathbb{%
O},\bullet ,n\right) $ has been discussed above (within the treatment of $p\mathbb{O}$ itself), whereas the transition\ $%
\left( \mathbb{O},\bullet ,n\right) \longrightarrow \left( \mathbb{O},\cdot
,n\right) $ can be realized through the aid of the para-unit $\boldsymbol{1}\in p\mathbb{O%
}$, such that
\begin{equation}
x\cdot y=\left( \boldsymbol{1}\bullet x\right) \bullet \left( y\bullet \boldsymbol{1}\right) =%
\overline{x}\bullet \overline{y}.
\end{equation}%
Again, the new algebra $\left( p\mathbb{O},\cdot ,n\right) $ is a $8$%
-dimensional composition algebra which is also unital and division; thus,
for the Hurwitz theorem, it must necessarily be isomorphic to the octonions
$\mathbb{O}$ : $\left( p\mathbb{O},\cdot ,n\right) \simeq \mathbb{O}$.

Finally, since $\tau \left( \overline{x}\right) =\overline{\tau \left(
x\right) }$, the transition $\left( \mathbb{O},\bullet ,n\right)
\longrightarrow \left( \mathbb{O},\ast ,n\right) $ is realized by defining
the Okubonic product $\ast $ in terms of the para-octonionic product, by
exploiting the order-3 involution $\tau $ :
\begin{equation}
x\ast y=\tau \left( x\right) \bullet \tau ^{2}\left( y\right) ,
\end{equation}%
and reversely the transition $\left( \mathbb{O},\ast ,n\right)
\longrightarrow \left( \mathbb{O},\bullet ,n\right) $ can be realized by
defining the para-octonionic product $\bullet $ in terms of the Okubonic
one, once again by using $\tau $ :%
\begin{equation}
x\bullet y=\tau ^{2}\left( x\right) \ast \tau \left( y\right) .
\end{equation}

Thus, we have shown how all the three 8-dimensional, real, division composition
algebras are obtainable one from the other (see Table \ref%
{tab:Oku-Para-Octo} for a summary).
\begin{table}
\centering{}%
\begin{tabular}{|c|c|c|c|}
\hline
Algebra & $\mathcal{O}\equiv \left( \mathbb{O},\ast ,n\right)$ & $p\mathbb{O}\equiv \left( \mathbb{O},\bullet ,n\right)$ & $\mathbb{O}\equiv \left( \mathbb{O},\cdot ,n\right)$\tabularnewline
\hline
\hline
$x*y$ & $x*y$ & $\tau\left(x\right)\bullet\tau^{2}\left(y\right)$ & $\tau\left(\overline{x}\right)\cdot\tau^{2}\left(\overline{y}\right)$\tabularnewline
\hline
$x\bullet y$ & $\tau^{2}\left(x\right)*\tau\left(y\right)$ & $x\bullet y$ & $\overline{x}\cdot\overline{y}$\tabularnewline
\hline
$x\cdot y$ & $\left(e*x\right)*\left(y*e\right)$ & $\left(\boldsymbol{1}\bullet x\right)\bullet\left(y\bullet\boldsymbol{1}\right)$ & $x\cdot y$\tabularnewline
\hline
\end{tabular}\caption{\label{tab:Oku-Para-Octo}In this table we see how to obtain the Okubonic
product $*,$ the para-octonionic product $\bullet$ and the octonionic
product $\cdot$ from Okubo algebra $\left(\mathbb{O},*,n\right)$,
para-octonions $\left(p\mathbb{O},\bullet,n\right)$ and octonions
$\left(\mathbb{O},\cdot,n\right)$, respectively.}
\end{table}
However, it is worth remarking once again that these algebras are not
isomorphic.

\subsection{Symmetries}

%Symmetries of the composition algebras introduced above will play a
%significant role in our subsequent treatment.

Given an algebra $\left( A,\cdot ,n\right) $, one can consider :

\begin{itemize}
\item The special orthogonal group $\text{SO}\left( A\right) $, defined as
the identity connected component of the group of endomorphisms of $A$ that
preserve the norm $n$ (or equivalently the bilinear and symmetric inner product defined through its polarization), but not
necessarily the algebraic structure defined by the product\footnote{%
The upperscript $0$ denotes the identity connected component of the Lie
group.} $\cdot $ :%
\begin{equation}
\text{SO}\left( A\right) :=\left\{ \varphi \in \text{End}^{0}\left( A\right)
:n\left( \varphi \left( x\right) \right) =n\left( x\right) \right\} .
\end{equation}%
As usual, in the subsequent treatment $\text{Spin}\left( A\right) $ will denote the spin
covering group of $\text{SO}\left( A\right)$.

\item The automorphism group $\text{Aut}\left( A\right) $ is the group of
(linear, identity connected) isomorphisms that preserves the algebraic structure :%
\begin{equation}
\text{Aut}\left( A\right) :=\left\{ \varphi \in \text{End}^{0}\left( A\right)
:\varphi \text{\text{ bijective,~linear,~}}\varphi \left( x\cdot y\right)
=\varphi \left( x\right) \cdot \varphi (y)\right\} .
\end{equation}%
While an element of Aut$\left( A\right) $ is also an element of SO$\left(
A\right) $, the converse is not necessarily true. In fact, in general Aut$(A)
$ is a proper subgroup of SO$(A)$:%
\begin{equation}
\text{Aut}\left( A\right) \varsubsetneq \text{SO}\left( A\right) .
\end{equation}

\item The triality group is the group of triples of orthogonal
transformations of $A$ which respect its algebraic structure:
\end{itemize}

\begin{equation}
\text{Tri}\left( A\right) :=\left\{ \left( \alpha ,\beta ,\gamma \right) \in
\text{SO}\left( A\right) ^{\otimes 3}:\alpha \left( x\cdot y\right) =\beta
\left( x\right) \cdot \gamma \left( y\right) \right\} .
\end{equation}%
In general, Tri$\left( A\right) $ is a (proper) subgroup of the product
group SO$\left( A\right) ^{\otimes 3}$, and it contains (or coincides with)
SO$\left( A\right) $ itself :%
\begin{equation}
\text{SO}\left( A\right) \subseteq \text{Tri}\left( A\right) \varsubsetneq
\text{SO}\left( A\right) ^{\otimes 3}.
\end{equation}

The symmetries of $\mathbb{O}$, $p\mathbb{O}$ and $\mathcal{O}$ are
well-known (e.g., see \cite{El15,El21,Knus,SalzBGHS}) and they are
summarized in Table \ref{tab:Symmetries-of-the}.
\begin{table}
\begin{centering}
\begin{tabular}{|c|c|c|c|}
\hline
 & Octonions $\mathbb{O}$ & Para-Octonions $p\mathbb{O}$ & Okubo $\mathcal{O}$\tabularnewline
\hline
\hline
$\text{Aut}\left(A\right)$ & $\text{G}_{2(-14)}$ & $\text{G}_{2(-14)}$ & $\text{SU(3)}$\tabularnewline
\hline
$\text{SO}\left(A\right)$ & $\text{Spin}\left(8\right)$ & $\text{Spin}\left(8\right)$ & $\text{Spin}\left(8\right)$\tabularnewline
\hline
$\text{Tri}\left(A\right)$ & $\text{Spin}\left(8\right)$ & $\text{Spin}\left(8\right)$ & $\text{Spin}\left(8\right)$\tabularnewline
\hline
\end{tabular}\caption{\label{tab:Symmetries-of-the}Symmetries of the octonions $\mathbb{O}$,
para-octonions $p\mathbb{O}$ and Okubo algebra $\mathcal{O}$. }
\par\end{centering}
\end{table}
Both the appearance of the spin covering group Spin$\left( 8\right) $
(instead of SO$\left( 8\right) $) and the fact that Tri$\left( A\right)
\simeq $Spin$(8)$ are due to the triality of $\mathfrak{d}_{4}$ \cite{SalzBGHS, Albert,
Elduque-Okubo}.

\section{\label{Octonions}$\left( \mathbb{O},\cdot ,n\right) $ and QCD}

\subsection{\label{Symmetries-O} Symmetries}

Tautologically, any algebra $A$ sits in a linear (not necessarily irreducible)
representation of its (linearly realized) automorphism Lie group, whose Lie
algebra is the algebra of derivations, i.e. $\mathfrak{der}\left( A\right) =%
\mathfrak{Lie}\left( \text{Aut}\left( A\right) \right) $. Considering $%
\left( \mathbb{O},\cdot ,n\right) $, Table \ref{tab:Symmetries-of-the}
reports that
\begin{eqnarray}
\text{Aut}\left( \mathbb{O}\right) &=&\text{G}_{2(-14)}, \\
\text{Tri}\left( \mathbb{O}\right) &=&\text{Spin}\left( 8\right) , \\
\text{Spin}\left( \mathbb{O}\right) &=&\text{Spin}\left( 8\right) ,
\end{eqnarray}%
and it holds that
\begin{equation}
\mathbb{O}\simeq \mathbf{1}\oplus \mathbf{7}~\text{of~}\text{G}_{2(-14)}.
\end{equation}%
Both Spin$\left( 8\right) $ and $\text{G}_{2(-14)}$ have real
representations \cite{BK}. The relation among these Lie groups is given by
the following chain of maximal group embeddings :
\begin{equation}
\begin{array}{ccccc}
\text{Spin}\left( 8\right) & \supset _{\text{s}} & \text{Spin}\left( 7\right)
& \supset _{\text{ns}} & \text{G}_{2(-14)} \\
\mathbf{8}_{v} & = & \mathbf{7}\oplus \mathbf{1} & = & \mathbf{7}\oplus
\mathbf{1} \\
\mathbf{8}_{s} & = & \mathbf{8} & = & \mathbf{7}\oplus \mathbf{1} \\
\mathbf{8}_{c} & = & \mathbf{8} & = & \mathbf{7}\oplus \mathbf{1} \\
\mathbf{28} & = & \mathbf{21}\oplus \mathbf{7} & = & \mathbf{14}\oplus
\mathbf{7}\oplus \mathbf{7},%
\end{array}
\label{pre-Dynkinn}
\end{equation}%
where the subscripts \textquotedblleft $s$\textquotedblright\ and
\textquotedblleft $ns$\textquotedblright\ of $\subset $ respectively stand
for \textquotedblleft symmetric\textquotedblright\ and \textquotedblleft
non-symmetric\textquotedblright\ (embedding), and, as mentioned at the end
of the Introduction, we recall that the physicists' notation of
representations of Lie algebras and groups is used throughout this paper. In
particular, $\mathbf{8}_{v}$, $\mathbf{8}_{s}$ and $\mathbf{8}_{c}$ are the
three eight-dimensional irreducible representations of $Spin(8)$, namely the
vector, (semi)spinor and \ conjugate (semi)spinor, respectively, whereas $%
\mathbf{28}$ denotes the adjoint irrepr. of $Spin(8)$ itself.

We remark that the unital nature of $\mathbb{O}
$ is related to the fact that Aut$\left( \mathbb{O}\right) $ is a (maximal)
subgroup of Spin$\left( 7\right) $, and that one (say%
\footnote{%
This depends on the choice of \textquotedblleft spinor
polarization\textquotedblright\ of the Spin$(7)$ subalgebra \cite{Minchenko}%
, but it is immaterial, due to $\mathfrak{d}_{4}$ triality symmetry.}, $\mathbf{8}_{v}$%
) of the three 8-dimensional representations of Spin$\left( 8\right) $ contains a
singlet (namely, the unit $1$ of $\mathbb{O}$) when branched with respect to Spin$\left(
7\right) $; in fact, (\ref{pre-Dynkinn}) defines a choice of \textquotedblleft
polarization\textquotedblright\
within the triality symmetry of Spin$\left( 8\right) $ (see e.g. \cite%
{Porteous}). The existence of a unity implies the following decomposition of
$\mathbb{O}$ :
\begin{equation}
\mathbb{O}=\mathbb{R}\oplus \mathbb{O}^{\prime },
\end{equation}%
where $\mathbb{O}^{\prime }:=\text{span}_{\mathbb{R}}\left(e_{1},...,e_{7}\right) $
denotes the \textit{imaginary} octonions, whose basis
is provided by the seven octonionic units $e_{1},...,e_{7}$,
with multiplication rules given by a suitable realization of the discrete
geometric of the Fano plane, as shown in Fig. \ref{fig:Fano-Plane}.

\subsection{\label{O-Breakings} $\text{SU$\left( 3\right) $}_{\mathbb{O}}$ $%
\varsubsetneq $Aut$\left( \mathbb{O}\right) $}

The automorphism group of the octonions $\text{Aut}\left( \mathbb{O}\right) $
is the real compact form of the exceptional Lie group $\text{G}_{2}$, i.e.,
Aut$\left( \mathbb{O}\right) =$G$_{2(-14)}$. The Lie group $\text{G}%
_{2(-14)} $ admits three maximal subgroups, two of Borel-de Siebenthal type,
i.e., of maximal rank \cite{BdS}, namely $\text{SU}\left( 3\right) $ and SO$%
\left( 4\right) \simeq \text{SU}\left( 2\right) \times \text{SU}\left(
2\right) $, and one of non-maximal rank, $\text{SU}\left( 2\right) $. In the
present treatment, we are interested in the\footnote{%
By $\text{SU}(3)$ we denote the compact real form of the Lie group whose Lie
algebra is $\mathfrak{a}_{2}$, which is also the maximal compact subgroup of
the Lie group $\text{SL}(3,\mathbb{C})_{\mathbb{R}}$.} $\text{SU}\left(
3\right) $ maximal subgroup of $\text{G}_{2(-14)}$, which we will be
denoting as $\text{SU$\left( 3\right) $}_{\mathbb{O}}$, and in its
representations stemming from the decomposition of the defining and adjoint
irreprs. of G$_{2(-14)}$ itself :

\begin{equation}
\begin{array}{ccc}
\text{G}_{2(-14)} & \supset _{\text{ns}} & \text{SU$\left( 3\right) $}_{%
\mathbb{O}} \\
\mathbf{7} & = & \mathbf{3}\oplus \overline{\mathbf{3}}\oplus \mathbf{1,} \\
\mathbf{14} & = & \mathbf{8}\oplus \mathbf{3}\oplus \overline{\mathbf{3}}.%
\end{array}
\label{thisthis}
\end{equation}%
Despite the fact that all representations of $\text{G}_{2(-14)}$ are real, $%
\text{SU$\left( 3\right) $}$ (and thus $\text{SU$\left( 3\right) $}_{\mathbb{%
O}}$ defined in (\ref{thisthis})) admits complex (i.e., reflexive)
representations \cite{BK}. In the case under consideration, this implies the
existence of a complex structure $J$ in the exceptional (and non-symmetric)
presentation $S_{J}^{6}$ of the $6$-sphere with isotropy group $\text{SU$%
\left( 3\right) $}_{\mathbb{O}}$ (see e.g. \cite{Draper} and Refs. therein),
\begin{equation}
S_{J}^{6}\simeq \frac{\text{G}_{2(-14)}}{\text{SU$\left( 3\right) $}_{\mathbb{O}}}.
\label{SJ6}
\end{equation}%
Thus, Aut$\left( \mathbb{O}\right) $ can also be presented as
\begin{equation}
\text{G}_{2(-14)}\simeq \text{SU$\left( 3\right) $}_{\mathbb{O}}\ltimes
S_{J}^{6}.
\end{equation}%
There are seven maximal non-symmetric subgroups $\text{SU$\left( 3\right) $}%
_{\mathbb{O}}$ of $\text{G}_{2(-14)}$, defined as invariance symmetries of
one of the seven octonionic imaginary units; they are all equivalent under
inner automorphisms of $\text{G}_{2(-14)}$ itself, and they all correspond
to the following $\text{SU$\left( 3\right) $}_{\mathbb{O}}$-covariant
breaking of $\mathbb{O}$ :
\begin{equation}
\begin{array}{ccc}
\text{G}_{2(-14)} & \supset & \text{SU$\left( 3\right) $}_{\mathbb{O}} \\
\mathbf{1}\oplus \mathbf{7} & = & \mathbf{1}\oplus \mathbf{1}\oplus \mathbf{3%
}\oplus \overline{\mathbf{3}}, \\
\mathbb{O}=\mathbb{R}\oplus \mathbb{O}^{\prime } & = & \mathbb{C}\oplus
\mathbb{C}^{3},%
\end{array}
\label{this}
\end{equation}%
where one can identify
\begin{equation}
\mathbb{C}\simeq \mathbb{R}\oplus J\mathbb{R},  \label{C}
\end{equation}%
with $J$ denoting an arbitrary but fixed octonionic imaginary unit, i.e., $%
J\in \left\{e_{1},...,e_{7}\right\} $, which is $\text{SU$%
\left( 3\right) $}_{\mathbb{O}}$-invariant.

\subsection{Application to QCD}

The physical interpretation of the maximal subgroup $SU(3)_{%
\mathbb{O}}$ as the \textit{color} gauge group $SU(3)_{\text{color}}$ of the
strong interaction, i.e.,%
\begin{equation}
SU(3)_{\text{color}}\equiv SU(3)_{\mathbb{O}},  \label{Phys-O}
\end{equation}%
has a long history, dating back to Gürsey, Günaydin, Ramond, Sikivie
\cite{GG1}-\nocite{Gursey,GG2}\cite{GRS} and Morita \cite{M1}-\nocite{M2}\cite{M3} between mid '70s and early
'80s, and more recently revived by Bisht, Chanyal \textit{et al.} \cite{C1,C2,C3} and
Wolk \cite{Wolk}. In a broader framework, the algebraic formulation of the
particle content and the gauge symmetries of the SM and beyond, crucially
involving the octonions and related (exceptional) algebraic and geometric
structures, has been the object of a number of works along the years;
without any claim of completeness of our list, here we confine ourselves to
cite Dixon \cite{Dixon}, Furey \cite{F1}, Hughes \cite{FH1,FH2}, Dray,
Manogue and Wilson \cite{DMW1,DMW2,DMW3}, Boyle \cite{Boyle}, Krasnov \cite%
{Krasnov}, Todorov and Dubois-Violette \cite{TDV1}, Singh \cite%
{Singh1,Singh2}, Penrose \cite{Penrose}, Castro \cite{Castro}, Rowlands \cite%
{Rowlands}, Masi \cite{Masi}, and two of the present authors with Chester,
Aschheim and Irwin \cite{DR1}.

By virtue of (\ref{this}), the identification (\ref{Phys-O})
implies $\mathbb{O}$ to correspond to a color triplet (quark $q$) and its
conjugate anti-color anti-triplet (anti-quark $\overline{q}$), plus two
color singlets (whose possible identification will not be discussed here). Thus,
the decomposition (\ref{this}) can be interpreted in QCD as follows :
\begin{equation}
\begin{array}{ccc}
\text{G}_{2(-14)} & \supset  & \text{SU$\left( 3\right) $}_{\mathbb{O}} \\
\mathbb{O}=\mathbf{1}\oplus \mathbf{7} & = & \underset{\text{%
pair~of~QCD~singlets}}{\underbrace{\mathbf{1}\oplus \mathbf{1}}}\oplus
\underset{q}{\mathbf{3}}\oplus \underset{\overline{q}}{\overline{\mathbf{3}}}%
,%
\end{array}
\label{q}
\end{equation}%
with $q$ being any of the six quark flavours. Therefore, within
this interpretation, a pair of Tri$(\mathbb{O})\simeq $Spin$\left( 8\right) $%
-covariant (and $\mathfrak{d}_{4}$-triality-invariant) triplets of reciprocally
independent octonions contains the color states of \textit{all} quarks and
antiquarks (namely, a $\mathbf{3}\oplus \overline{\mathbf{3}}$ for every
quark flavour $q=u,d,c,s,t,b$) of the SM, plus a total of 12 color singlets%
\footnote{%
We cannot help but observe that this formula seems to provide further
evidence for the relevance of $\mathfrak{d}_{4}$-triality for the existence of three
generations of matter fields in the SM, an intriguing research venue which
has a quite long story \cite{Silagadze, Rubenthaler, Ito}. } :
\begin{equation}
\left\{ \mathbb{O}_{Av},\mathbb{O}_{As},\mathbb{O}_{Ac}\right\} _{A=1,2}=%
\underset{6\cdot \left( \mathbf{3}\oplus \overline{\mathbf{3}}\right) =\text{%
quark~\&~antiquark~color~states}}{\left\{ u,\overline{u},d,\overline{d},c,%
\overline{c},s,\overline{s},t,\overline{t},b,\overline{b},\right\} }\oplus
\underset{12~\text{color~singlets}}{12\cdot \mathbf{1}}.
\end{equation}%

On the other hand, the decomposition of the adjoint representation under the
same embedding as (\ref{q}), given by the second line of (\ref{thisthis}),
suggests that the gauge group $\text{SU$\left( 3\right) $}_{\text{color}}$%
, within the interpretation (\ref{Phys-O}), could be enhanced to the smallest
exceptional Lie group, $\text{G}_{2(-14)}$. This is something that
physicists have tried for a long time, in their quest for a comprehensive
Grand Unified Theory (GUT), finding an insurmountable obstruction in the
fact that, as mentioned above, $\text{G}_{2(-14)}$ has no complex
representations, and thus cannot properly account for the SM fermions. Thus,
despite recent attempts (see e.g. \cite{Masi}, and Refs. therein), no consistent GUT
with gauge group $\text{G}_{2(-14)}$ can be formulated.

Nevertheless, it is instructive, also in view of the treatment below, to
report the pattern of strong charges for

\begin{itemize}
\item the three colors of a quark $q$ in the irrepr. $\mathbf{3}$ of $\text{%
SU$\left(3\right)$}_{\text{color}}$;

\item three anti-colors of the corresponding anti-quark $\overline{q}$ in
the corresponding, conjugate irrepr. $\overline{\mathbf{3}}$ of $\text{SU$%
\left(3\right)$}_{\text{color}}$;

\item eight gluons in the adjoint irrepr. $\mathbf{Adj}\equiv\mathbf{8}$ of $%
\text{SU$\left(3\right)$}_{\text{color}}$.
\end{itemize}

This pattern is reported in the left side of Fig. \ref%
{fig:Strong-force-Magic-Star} with the two gluons $g^{3}$ and $g^{8}$ of
zero color charge overlapping in the center of the pattern, and
corresponding to the Cartan generators of $\mathfrak{g}_{2(-14)}$ or,
equivalently, of $\mathfrak{su}(3)$. In this pattern, of course $q$ can take
any value $q\in \left\{ u,d,c,s,t,b\right\} $.
\begin{figure}[h]
\centering{}\includegraphics[scale=0.8]{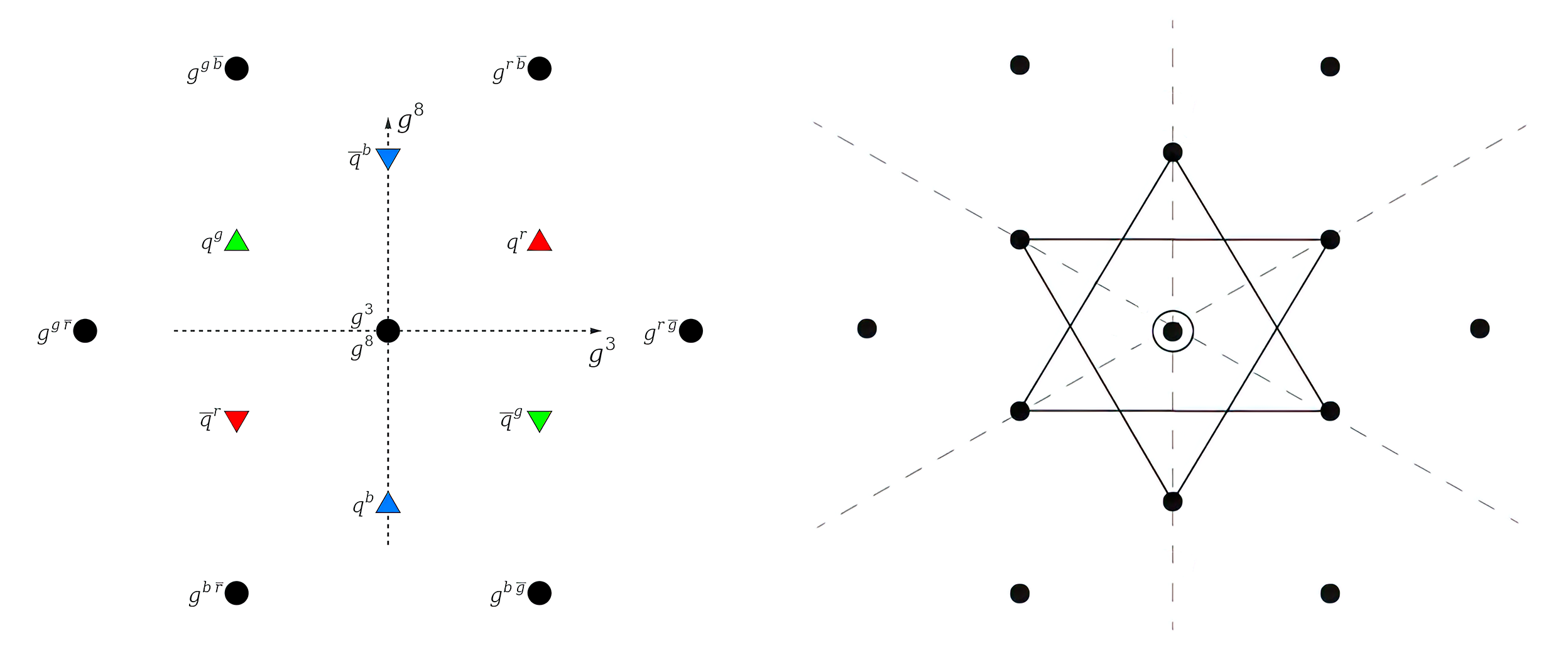}
\caption{\emph{On the left} : strong force symmetry pattern, pertaining to
the $\boldsymbol{3}$ color states of a quark $q$, and the corresponding $%
\overline{\boldsymbol{3}}$ states of its anti-quark $\overline{q}$, along
with the $\boldsymbol{8}$ gluon color states (with $g^{3}$ and $g^{8}$
coinciding in the center of the diagram, taken from \protect\cite{Lisi}).
\emph{On the right} : \textquotedblleft Magic Star\textquotedblright\
projection of the root lattice of $\mathfrak{g}_{2(-14)}$ \protect\cite%
{Mukai,MagicStar}, with the plane of the sheet being defined by the two
Cartan generators of $\mathfrak{g}_{2(-14)}$ itself. Despite the fact that
the Cartans of $\mathfrak{g}_{2(-14)}$ (on the right) and of $\mathfrak{su}%
(3)$ (on the left) coincide, the strong force pattern in the l.h.s. cannot
be interpreted as the \textquotedblleft Magic Star\textquotedblright\
projection of $\mathfrak{g}_{2(-14)}$, and vice versa. This is ultimately
due to the fact that the Lie groups SU$\left( 3\right) _{\mathcal{O}}$
(pertaining to the l.h.s) and SU$\left(
3\right) _{\mathbb{O}}$ (pertaining to the r.h.s.) are totally different
subgroups of SO$\left( \mathbb{O}\right) \simeq $Spin$\left( 8\right) \simeq
$SO$\left( \mathcal{O}\right) $; see (\ref{diff}), and the discussion at the
end of Sec. 4.}
\label{fig:Strong-force-Magic-Star}
\end{figure}

\section{\label{Okubonions}$\mathcal{O}$ and QCD}

\subsection{Symmetries}

As reported in Table \ref{tab:Symmetries-of-the}, the Okubo algebra $\mathcal{O}\equiv \left( \mathbb{O},\ast
,n\right) $  is characterized by the following
symmetries :
\begin{eqnarray}
\text{Aut}\left( \mathcal{O}\right) &=&\text{\text{SU}$\left( 3\right)
\equiv $SU}\left( 3\right) _{\mathcal{O}},  \label{Aut-Ok} \\
\text{Tri}\left( \mathcal{O}\right) &=&\text{Spin}\left( 8\right) , \\
\text{Spin}\left( \mathcal{O}\right) &=&\text{Spin}\left( 8\right) ,
\end{eqnarray}%
where we introduced $\text{SU}\left( 3\right) _{\mathcal{O}}$ as the
automorphism group $\text{Aut}\left( \mathcal{O}\right) $, in order to
distinguish it from the $\text{SU}\left( 3\right) _{\mathbb{O}}$ introduced
in (\ref{thisthis}). It also holds that
\begin{equation}
\mathcal{O}\simeq \mathbf{8}\equiv \mathbf{Adj}~\text{of~}\text{SU}\left(
3\right) _{\mathcal{O}}.  \label{4}
\end{equation}

\subsection{$\text{SU}\left( 3\right) _{\mathcal{O}}=\text{Aut}(\mathcal{O})$}

Spin$\left( 8\right) $ and\footnote{$SU(3)_{\mathcal{O}}$
admits two maximal subgroups, one of Borel - de Siebenthal (i.e., maximal rank) type,
i.e. U$(2)$, and one of non-maximal rank, i.e. SU$(2)$. Their treatment will
be considered elsewhere.} $\text{SU}\left( 3\right) _{\mathcal{O}}$ are related by a
maximal (and non-symmetric) embedding, under which the defining and
the adjoint irrepr. of Spin$\left( 8\right) $ decompose as follows :
\begin{equation}
\begin{array}{ccc}
\text{Spin}\left( 8\right) & \supset _{\text{ns}} & \text{SU}\left( 3\right)
_{\mathcal{O}} \\
\mathbf{8}_{v} & = & \mathbf{8} \\
\mathbf{8}_{s} & = & \mathbf{8} \\
\mathbf{8}_{c} & = & \mathbf{8} \\
\mathbf{28} & = & \mathbf{8}\oplus \mathbf{10}\oplus \overline{\mathbf{10}},%
\end{array}
\label{Dynkinn}
\end{equation}%
where $\mathbf{10\equiv }S^{3}\mathbf{3}$ is the rank-3 symmetric irrepr. of
$\text{SU}\left( 3\right) _{\mathcal{O}}$ . The maximal embedding (\ref%
{Dynkinn}) exists by virtue of a Theorem of Dynkin (see e.g. Th. 1.5 of \cite%
{Dynkin}), and it is a consequence of the existence of the Cartan-Killing
bilinear invariant form in the adjoint irrepr. $\mathbf{8}$
 of the Lie group $\text{SU}\left( 3\right) _{\mathcal{O}}$ .

We stress that the non-unital nature of $\mathcal{O}$ is deeply linked to the fact that Aut$%
\left( \mathcal{O}\right) $ is a \textit{maximal} subgroup of Spin$\left(
8\right) $, and \textit{not} a subgroup of Spin$\left( 7\right) $. The lack
of a unity implies that $\mathcal{O}$ is \textit{irreducible} under the
action of its automorphism group $\text{SU}\left( 3\right) _{\mathcal{O}}$ ,
as expressed by (\ref{4}). Thus, differently from the (unique) unit element $1$ in any unital algebra (which
is invariant under the automorphism group of the unital algebra itself), in the basis of the Okubonic 8-dimensional vector space (whose a possible explicit realization is given by(\ref{eq:definizione i ottonioniche})),
the (not unique, but rather threefold\footnote{%
From the very definition of $\mathcal{O}$ as the algebra of $3\times 3$
traceless Hermitian matrices over $\mathbb{C}$ (namely, as $\mathfrak{J}%
_{3}\left( \mathbb{C}\right) _{0}$) with a suitable deformation $\ast $ of
the Michel-Radicati product \cite{Ok2}, it follows that there are three
independent idempotents in $\mathcal{O}$, which are however all equivalent
under the order -three involutive automorphism $\tau $ of $\mathcal{O}$ (cfr. (1.5))
\cite{CorMarZuc,El15}.}) idempotent $e$ is \textit{not} invariant under Aut$\left( \mathcal{O}\right) $, but instead it
corresponds to one generator inside the adjoint representation of Aut$\left(
\mathcal{O}\right) $ itself.

Before discussing the possible relevance of Okubonions in QCD, we would like
to comment that, intuitively, a unital algebra (such as $\mathbb{O}\equiv
\left( \mathbb{O},\cdot ,n\right) $) might turn out to be more amenable at
describing theories and phenomena which admit a perturbative description,
whereas a non-unital algebra (such as $\mathcal{O}\equiv \left( \mathbb{O}%
,\ast ,n\right) $) might result in a more consistent algebraic modeling of
non-perturbative regimes and physical phenomena. For this reason, we will
not be dealing with the seamless unital extension of $\mathcal{O}$, defined
by $\mathcal{O}^{+}:=\mathcal{O}\oplus 1$, since this is spoiled of the
peculiar property of non-unitality, and, for instance, due to the traceful
nature of the unit element, it does \textit{not} admit a realization as $%
\mathfrak{J}_{3}(\mathbb{C})_{0}$ (with suitable deformation of the Jordan
product \textit{à la Michel-Radicati}) anymore.

\subsection{Application to QCD}

Hinted by (\ref{Aut-Ok}), we cannot help but put forward a conjectural
physical interpretation of the Okubonic automorphism group $\text{SU}\left(
3\right) _{\mathcal{O}}$ (whose relation with (\ref{Phys-O}) will be
investigated below) as the \textit{color} gauge group $\text{SU$\left(
3\right) $}_{\text{color}}$ of the strong interaction within the SM, i.e.,
\begin{equation}
\text{SU$\left( 3\right) $}_{\text{color}}\equiv \text{SU}\left( 3\right) _{%
\mathcal{O}}.  \label{Phys-Ok}
\end{equation}%
By virtue of (\ref{4}), (\ref{Phys-Ok}) implies $\mathcal{O}$ to represent
the octet of the \textit{gluons }$\left\{ g^{i}\right\} _{i=1,...,8}$,
namely of the gauge vector bosons mediating the strong interaction\footnote{%
It is here worth remarking that, within the physical interpretation (\ref%
{Phys-Ok}), any $\text{SU}\left( 2\right) $ or $\text{U}\left( 1\right) $
subgroup of $\text{SU}\left( 3\right) _{\mathcal{O}}$ \textit{cannot} be
regarded as the weak or electromagnetic gauge group, of course.} in QCD :
\begin{equation}
\mathcal{O}\equiv \left( \mathbb{O},\ast ,n\right)\simeq \underset{\left\{ g^{i}\right\} _{i=1,...,8}}{\mathbf{8}}%
\equiv \mathbf{Adj}~\text{of~}\text{SU}\left( 3\right) _{\mathcal{O}}.
\label{gluons}
\end{equation}%
Thus, there is a $1:1$ correspondence between the elements of the basis of the Okubonic 8-dimensional vector space (whose a possible explicit realization is given by(\ref{eq:definizione i ottonioniche})) and the gluons $\left\{ g_{i}\right\} _{i=1,...,8}$ of QCD;
this is consistent with the physical picture in which, since the color
symmetry (\ref{Phys-Ok}) is exact and unbroken, \textit{all gluons are
massless and stand on the same footing, color-wise}. The gluons are indeed
in $1:1$ correspondence with the generators of $\text{SU}\left( 3\right) $
or, realization-wise, with the eight Gell-Mann matrices $\lambda ^{i}$'s
(see e.g. \cite{Griffiths}, pp. 283-288 and 366-369 therein).

However, it is at this stage worth stressing the conjectural nature of the identification (\ref{Phys-Ok}),
which deserves further work and investigation. Nonetheless, it is suggestive
to wonder whether any feature of QCD, such as asymptotic freedom and
color confinement, might be traced back to the unusual properties of $%
\mathcal{O}$ itself, such as the non-alternativity and the absence of a unit
element (i.e., non-unitality), despite $\mathcal{O}$ is still a division algebra.

\section{$\text{SU}\left(3\right)_{\mathcal{O}}$ and $\text{SU}%
\left(3\right)_{\mathbb{O}}$ as different subgroups of Spin$\left(8\right)$}

We have put forward a suggestive physical interpretation of the (real)
Okubo algebra $\mathcal{O}$ in (an algebraic model of) QCD, given by (\ref%
{gluons}) (within the identification (\ref{Phys-Ok})). By recalling its
octonionic counterparts, i.e. (\ref{q}) (within the identification (\ref%
{Phys-O})), it then seems that $\mathbb{O}$ and $\mathcal{O}$ have
complementary\footnote{%
Such a `physical complementarity', corresponding to switching between
the matter and the gauge sectors of QCD, is actually realized by the
definition of alternative product within the vector space of the algebra
(while keeping the norm $n$ invariant); in fact, from its very definition, $%
\mathcal{O}\equiv \left( \mathbb{O},\ast ,n\right) $ (see Sec. 1), and $%
\mathbb{O}\equiv \left( \mathcal{O},\cdot ,n\right) $ (cf. the discussion
below (\ref{tthis})).} physical interpretations in QCD (namely, quarks/fermions from $\mathbb{O}$
 \textit{versus} gluons/bosons from $\mathcal{O}$).

In order to elucidate the relation between the physical identifications (\ref{Phys-O})
and (\ref{Phys-Ok}), in this Section we will show that the two corresponding SU$(3)$,
namely $\text{SU}\left( 3\right) _{\mathbb{O}}$ and $\text{SU}\left(
3\right) _{\mathcal{O}}$, do \textit{not} coincide, but they are rather
totally different. This is a consequence of the different embeddings of $%
\text{SU}\left( 3\right) _{\mathcal{O}}$ and $\text{SU}\left( 3\right) _{%
\mathbb{O}}$ in $\text{SO}\left( \mathbb{O}\right) \simeq \text{Spin}\left( 8\right) \simeq
\text{SO}\left( \mathcal{O}\right)$. Indeed, from (\ref{pre-Dynkinn}), (%
\ref{Dynkinn}) and (\ref{thisthis}), one can draw the following chains of
maximal group embeddings from Spin$(8)$ to $\text{SU}\left( 3\right) _{%
\mathcal{O}}$,
\begin{equation}
\begin{array}{ccc}
\text{Spin}\left( 8\right) & \supset & \text{SU}\left( 3\right) _{\mathcal{O}%
} \\
\mathbf{8}_{v} & = & \mathbf{8} \\
\mathbf{8}_{s} & = & \mathbf{8} \\
\mathbf{8}_{c} & = & \mathbf{8} \\
\mathbf{28} & = & \mathbf{8}\oplus \mathbf{10}\oplus \overline{\mathbf{10}},%
\end{array}
\label{uno}
\end{equation}%
and to $\text{SU}\left( 3\right) _{\mathbb{O}}$ ,
\begin{equation}
\begin{array}{ccccccc}
\text{Spin}\left( 8\right) & \supset & \text{Spin}\left( 7\right) & \supset
& G_{2(-14)} & \supset & \text{SU}\left( 3\right) _{\mathbb{O}} \\
\mathbf{8}_{v} & = & \mathbf{7}\oplus \mathbf{1} & = & \mathbf{7}\oplus
\mathbf{1} & = & \mathbf{3}\oplus \overline{\mathbf{3}}\oplus 2\cdot \mathbf{%
1} \\
\mathbf{8}_{s} & = & \mathbf{8} & = & \mathbf{7}\oplus \mathbf{1} & = &
\mathbf{3}\oplus \overline{\mathbf{3}}\oplus 2\cdot \mathbf{1} \\
\mathbf{8}_{c} & = & \mathbf{8} & = & \mathbf{7}\oplus \mathbf{1} & = &
\mathbf{3}\oplus \overline{\mathbf{3}}\oplus 2\cdot \mathbf{1} \\
\mathbf{28} & = & \mathbf{21}\oplus \mathbf{7} & = & \mathbf{14}\oplus
2\cdot \mathbf{7} & = & \mathbf{8}\oplus 3\cdot \mathbf{3}\oplus 3\cdot
\overline{\mathbf{3}}\oplus 2\cdot \mathbf{1}.%
\end{array}
\label{due}
\end{equation}%
Comparing the branchings of the $\mathbf{Adj}\equiv \mathbf{28}$ of Spin$(8)$
along the chains (\ref{uno}) and (\ref{due}), one immediately realizes that%
\begin{equation}
\text{SU}\left( 3\right) _{\mathcal{O}}\neq \text{SU}\left( 3\right) _{%
\mathbb{O}}.  \label{diff}
\end{equation}

Even more interestingly, $\text{SU}\left( 3\right) _{\mathcal{O}}$ and $%
\text{SU}\left( 3\right) _{\mathbb{O}}$ do \textit{not} even share a common $%
\text{SU}\left( 2\right) $ subgroup. This can be proved as follows. By
continuing the chain (\ref{uno}) to the maximal subgroups $\text{SU}\left(
2\right) \times \text{U}\left( 1\right) $ resp. $\text{SU}\left( 2\right) $
of $\text{SU}\left( 3\right) _{\mathcal{O}}$, one respectively obtains
\begin{eqnarray}
&&%
\begin{array}{l}
\text{Spin}\left( 8\right) \supset \text{SU}\left( 3\right) _{\mathcal{O}%
}\supset \text{SU}\left( 2\right) \times \text{U}\left( 1\right) , \\
\mathbf{28}=\mathbf{8}\oplus \mathbf{10}\oplus \overline{\mathbf{10}}=3\cdot
\mathbf{3}_{0}\oplus 2\cdot \mathbf{2}_{3}\oplus 2\cdot \mathbf{2}%
_{-3}\oplus \mathbf{4}_{3}\oplus \mathbf{4}_{-3}\oplus \mathbf{1}_{0}\oplus
\mathbf{1}_{-6}\oplus \mathbf{1}_{6}; \\
\\
\end{array}
\label{uno-1} \\
&&%
\begin{array}{l}
\text{Spin}\left( 8\right) \supset \text{SU}\left( 3\right) _{\mathcal{O}%
}\supset \text{SU}\left( 2\right) , \\
\mathbf{28}=\mathbf{8}\oplus \mathbf{10}\oplus \overline{\mathbf{10}}=3\cdot
\mathbf{3}\oplus \mathbf{5}\oplus 2\cdot \mathbf{7.}%
\end{array}
\label{uno-2}
\end{eqnarray}%
On the other hand, by continuing the chain (\ref{due}) to the maximal
subgroups $\text{SU}\left( 2\right) \times \text{U}\left( 1\right) $ resp. $%
\text{SU}\left( 2\right) $ of $\text{SU}\left( 3\right) _{\mathbb{O}}$ , one
respectively obtains
\begin{eqnarray}
&&%
\begin{array}{l}
\text{Spin}\left( 8\right) \supset ...\supset \text{SU}\left( 3\right) _{%
\mathbb{O}}\supset \text{SU}\left( 2\right) \times \text{U}\left( 1\right) ,
\\
\mathbf{28}=...=\mathbf{8}\oplus 3\cdot \mathbf{3}\oplus 3\cdot \overline{%
\mathbf{3}}\oplus 2\cdot \mathbf{1}=\mathbf{3}_{0}\oplus \mathbf{2}%
_{3}\oplus \mathbf{2}_{-3}\oplus 3\cdot \left( \mathbf{2}_{1}\oplus \mathbf{2%
}_{-1}\right) \oplus 3\cdot \left( \mathbf{1}_{0}\oplus \mathbf{1}%
_{-2}\oplus \mathbf{1}_{2}\right) ;%
\end{array}
\label{due-1} \\
&&%
\begin{array}{l}
\text{Spin}\left( 8\right) \supset ...\supset \text{SU}\left( 3\right) _{%
\mathbb{O}}\supset \text{SU}\left( 2\right) , \\
\mathbf{28}=...=\mathbf{8}\oplus 3\cdot \mathbf{3}\oplus 3\cdot \overline{%
\mathbf{3}}\oplus 2\cdot \mathbf{1}=7\cdot \mathbf{3}\oplus \mathbf{5}\oplus
2\cdot \mathbf{\mathbf{1}.}%
\end{array}
\label{due-2}
\end{eqnarray}%
The comparison of the final breakings (\ref{uno-1}),(\ref{uno-2}), (\ref%
{due-1}) and (\ref{due-2}) implies that $\text{SU}\left( 3\right) _{\mathcal{%
O}}$ and $\text{SU}\left( 3\right) _{\mathbb{O}}$, as subgroups of the same
Spin$(8)$ group, do \textit{not} even share a common $\text{SU}\left(
2\right) $ subgroup, \textit{q.e.d. }$ .\square \medskip $

Thus, (\ref{diff}) implies that \textit{no Okubonic/gluonic interpretation
of }$\mathfrak{su}(3)_{\mathbb{O}}$ (as found in (\ref{due})) is possible,%
\textit{\ }because $\text{SU}\left( 3\right) _{\mathcal{O}}$ and $\text{SU}%
\left( 3\right) _{\mathbb{O}}$ are different, inequivalent subgroups of Spin$%
\left( 8\right) $. This implies that the QCD pattern reported in the l.h.s.
of Fig. \ref{fig:Strong-force-Magic-Star}, which is intrinsically gluonic/Okubonic
(according to the interpretation (\ref{Phys-Ok}), \textit{cannot} enjoy an algebraic
interpretation as the \textquotedblleft Magic Star\textquotedblright\
decomposition of $\mathfrak{g}_{2(-14)}$, represented in the r.h.s. of
Fig. \ref{fig:Strong-force-Magic-Star}, which is instead intrinsically pertaining
to quarks/octonions (according to the interpretation (\ref{Phys-O})). This is ultimately due to (\ref{diff}). For the same reason, \textit{no
octonionic/quark} \textit{interpretation of }$\mathfrak{su}(3)_{\mathcal{O}}$
(as found in (\ref{uno})) is possible, of course.

\section{Conclusions}

In this work, for the first time to the best of our knowledge, a physical
interpretation of the real\footnote{%
The present paper only assumes $\mathbb{R}$ as ground field. We leave the
treatment on $\mathbb{C}$ and over finite fields $\mathbb{F}_{p}$ for future
investigation.}, non-unital, non-alternative, composition and division algebra of the Okubonions $%
\mathcal{O}\equiv \left( \mathbb{O},\ast ,n\right) $ has been put forward,
within (the attempts at elaborating an algebraic model of) QCD. As
investigated in \cite{EldMS1,EldMS2,CorMarZuc}, many algebraic and geometric
structures, such as the the octonionic entries of the Tits-Freudenthal Magic
Square and the Cayley plane, can be constructed by resorting to $\mathcal{O}$
only, without ever employing the algebra of the octonions.

Both $\mathbb{O}$ and $\mathcal{O}$ are non-associative, flexible,
composition and division, real 8-dimensional algebras. Such two algebras are
however very different: for instance, $\mathbb{O}$ is alternative and unital
(Hurwitz), whereas $\mathcal{O}$ is non-alternative and non-unital (only
admitting idempotent elements, though). Moreover, the Lie group Aut$\left(
\mathcal{O}\right) =\text{SU}\left( 3\right) _{\mathcal{O}}$ is much smaller
than (and not a subgroup, albeit with the same rank, of) the exceptional Lie
group $\text{G}_{2(-14)}=$Aut$(\mathbb{O})$. More precisely, Aut$\left( \mathcal{O}\right) =$SU$\left( 3\right) _{\mathcal{O}}$ and Aut$%
\left( \mathbb{O}\right) =$G$_{2(-14)}$ are totally different, rank-2
subgroups of SO$\left( \mathbb{O}\right) \simeq $Spin$\left( 8\right) \simeq
$SO$\left( \mathcal{O}\right) $ : SU$\left( 3\right) _{%
\mathcal{O}}$ is \textit{maximal} and non-symmetric in Spin$\left( 8\right) $
(as given by the embedding (\ref{uno})), whereas SU$\left( 3\right) _{%
\mathbb{O}}$ is \textit{next-to-maximal} (and non-symmetric) in Spin$\left(
8\right) $ (as given by (\ref{due})). We find this fact quite tantalizing,
and possibly paving the way to surprising developments.
Indeed, as shown in \cite{Elduque Gradings SC} and very
recently in \cite{CorMarZuc}, by using the Okubonions one can obtain a number of exceptional structures, such as
the cubic exceptional Jordan algebra $\mathfrak{J}_{3}(\mathbb{O})$ (\emph{%
aka} Albert algebra) or various real forms of the exceptional Lie groups of type $%
\text{F}_{4}$, $\text{E}_{6}$ and even $\text{E}_{7}$ and $\text{E}_{8}$,
without employing the octonions $\mathbb{O}$, nor $\text{G}_{2}$, at all.

Having proved that the $\text{SU}\left( 3\right) $ Lie groups pertaining to $%
\mathbb{O}$ and $\mathcal{O}$ are different, inequivalent subgroups of Spin$%
\left( 8\right) $, we hope to have been able in this paper to convey the
message that the uses of such composition algebras within the algebraic
modeling of the strong interaction in the SM are deeply different one from
the other, and, in a sense, complementary. In this regard, we would like to
remark that, in the perspective of an algebraic modeling of the strong
interaction, $\mathcal{O}$ may provide a more suitable algebraic structure,
with a remarkably smaller number of generators than $\mathbb{O}$ itself. Indeed, Aut$\left(
\mathcal{O}\right) =\text{SU}\left( 3\right) $, thus possibly matching the
exact unbroken gauge symmetry of QCD (within the conjectural interpretation (\ref{Phys-Ok})),
without the need to introduce further
(namely, six) massive gauge bosons corresponding to the generators of the
(non-symmetric) coset $\text{G}_{2\left( -14\right) }/\text{SU}\left(
3\right) $, as implied in $\text{G}_{2}$-based GUT extensions of the SM.

Moreover, it may well be that some `weird', unusual
properties (such as non-alternativity and non-unitality) of real Okubonions $%
\mathcal{O}$ might be related to peculiar QCD phenomena like
asymptotic freedom and color confinement (analogously to what has been discussed
for octonions in \cite{C3}), though the actual mechanisms remain to be
investigated.

Before concluding this work, we would like to stress once again the
conjectural nature of the identification (\ref{Phys-Ok}), which surely
deserves further work, e.g. in relation to the consistency of `Okubonic
fields', whose investigation seemingly calls for the Okubonic analogue of
the treatment given for the octonions in \cite{M1, M2, M3}. Also, at
present it is not known whether a representation of the Poincaré group can
be found within the Okubo algebra, and whether this is
consistent with the spin-1 (gluonic) interpretation put forward by (\ref%
{gluons}). We leave these intriguing issues for further future work.\medskip

We would like to conclude by citing Susumu Okubo in the introduction of his book
\cite{Ok-book}, dating back to 1995 : \textquotedblleft \textit{Octonion
algebra may surely be called a beautiful mathematical entity. Nevertheless,
it has never been systematically utilized in physics in any fundamental
fashion, although some attempts have been made toward this goal. However, it
is still possible that non-associative algebras (other than Lie algebras)
may play some essential future role in the ultimate theory, yet to be
discovered.}\textquotedblright\ While there have been many advances on the
physical applications of octonions since the publication of Okubo's book,
non-associative algebras may still be a key player in the formulation of a
ultimate theory of Physics. While octonions have been (with very good
reason) dubbed \textit{`the strangest numbers in string theory'} \cite%
{Baez-strange}, in the present paper we have suggested a fierce (possibly
even stranger!) opponent : the Okubonions.

\section*{Acknowledgments}

We would like to thank the anonymous referee, for her/his valuable remarks and constructive criticism,
which helped us to greatly improve the manuscript. The work of AM is supported by a \textquotedblleft
Maria Zambrano\textquotedblright{} distinguished researcher fellowship, financed by the European Union
within the NextGenerationEU program. This article is based upon work from COST Action CaLISTA CA21109 supported by COST (European Cooperation in Science and Technology).

$*$\noun{ Departamento de Matemática, }\\
\noun{Universidade do Algarve, }\\
\noun{Campus de Gambelas, }\\
\noun{8005-139 Faro, Portugal}
\begin{verbatim}
a55499@ualg.pt

\end{verbatim}
$\dagger$\noun{ Instituto de Física Teorica, Dep.to de Física,}\\
\noun{Universidad de Murcia, }\\
\noun{Campus de Espinardo, }\\
\noun{E-30100, Spain}
\begin{verbatim}
alessio.marrani@um.es

\end{verbatim}
$\ddagger$\noun{ Dipartimento di Scienze Matematiche, Informatiche
e Fisiche, }\\
\noun{Università di Udine, }\\
\noun{Udine, 33100, Italy}
\begin{verbatim}
francesco.zucconi@uniud.it
\end{verbatim}

\end{document}